\documentclass[twocolumn,showpacs,prl,nofootinbib]{revtex4}

\usepackage{graphicx}
\usepackage{dcolumn}
\usepackage{bm}
\usepackage{footnote}

\begin{document}
\renewcommand{\thefootnote}{\fnsymbol{footnote}}

\title{A model independent approach to the dark energy
equation of state}    

\author{P.S. Corasaniti}

\author{E.J. Copeland}

\affiliation{Centre for Theoretical Physics,
University of Sussex, Brighton, BN1 9QJ \\ United Kingdom }

\date{\today}

\begin{abstract}

The consensus of opinion in cosmology is that the Universe is
currently undergoing a period of accelerated expansion. With
current and proposed high precision experiments it offers the hope
of being able to discriminate
between the two competing models that are being suggested to
explain the observations, namely a cosmological constant or a time
dependent `Quintessence' model. The latter suffers from a plethora
of scalar field potentials all leading to similar late time
behaviour of the universe, hence to a lack of predictability. In
this paper, we develop a model independent approach which simply
involves parameterizing the dark energy equation of state in terms
of known observables. This allows us to analyse the impact dark energy
has had on cosmology without the need to refer to particular scalar
field models and opens up the possibility that future
experiments will be able to constrain the dark energy equation of
state in a model independent manner.
\end{abstract}

\maketitle

\section{Introduction}

Current cosmological observations suggest that the energy density
of the universe is dominated by an unknown type of matter, called
'dark energy', \cite{PERL,RIES,BOOM1,BOOM2,DASI,2df}. If correct,
it is characterized by a negative value for the equation of state
parameter $w$ (close to $-1$) and is responsible for the present
accelerating expansion of the Universe. The simplest dark energy
candidate is the cosmological constant, but in alternative
scenarios the accelerated phase is driven by a dynamical scalar
field called `Quintessence' \cite{WETT,RATRA,CALD,FERRE}. Although
recent analysis of the data provides no evidence for the need of a
quintessential type contribution \cite{CORAS,BEAN,HANN}, the
nature of the dark energy remains illusive. In fact the present
value of its equation of state $w_Q$ is constrained to be close to
the cosmological constant one, but the possibility of a time
dependence of $w_Q$ or a coupling with cold dark matter for
example \cite{LUCA} cannot be excluded. Recent studies have
analysed our ability to estimate $w_Q$ with high redshift object
observations \cite{WAGA,NEWMAN1,NEWMAN2,ALCANIZ,CALVAO} and to
reconstruct both the time evolution of $w_Q$
\cite{HUT00,ASTIER,BARG,GOL,MAO,WELLA,WELLB} as well as the scalar
field potential $V(Q)$ \cite{HUT99,SAINI,CHI}. An alternative
method for distinguishing different forms of dark energy has been
introduced in \cite{SAHNI}, but generally it appears that it will
be difficult to really detect such time variations of $w_Q$ even
with the proposed SNAP satellite \cite{KUJ,MAOR}. One of the
problems that has been pointed out is that, previously unphysical
fitting functions for the luminosity distance have been used,
making it difficult to accurately reproduce the properties of a
given quintessence model from a simulated data sample
\cite{HUT00,WELLB,GERKE}. A more efficient approach consists of
using a time expansion of $w_Q$ at low redshifts. For instance in
\cite{ASTIER,WELLA,WELLB} a polynomial fit in redshift space $z$
was proposed, while in \cite{EFS,GERKE} a logarithmic expansion in
$z$ was proposed to take into account a class of quintessence
models with slow variation in the equation of state. However
even these two expansions are limited, in that they can not
describe models with a rapid variation in the equation of state,
and the polynomial expansion introduces a number of unphysical
parameters whose value is not directly related to the properties
of a dark energy component.
The consequence is that their application is limited to low
redshift measurements and cannot be extended for example to the
analysis of CMB data. An interesting alternative to the fitting
expansion approach, has recently been proposed \cite{Huterer}, in
which the time behavior of the equation of state can be
reconstructed from cosmological distance measurements without
assuming the form of its parametrization. In spite of the
efficiency of such an approach, it does not take into account the
effects of the possible clustering properties of dark energy which
become manifest at higher redshifts. Hence its application has to
be limited to the effects dark energy can produce on the expansion
rate of the universe at low redshifts. On the other hand, it has
been argued that dark energy does not leave a detectable imprint
at higher redshifts, since it has only recently become the
dominant component of the universe. Such a statement, however, is
model dependent, on the face of it there is no reason why the dark
energy should be negligible deep in the matter dominated era. For
instance CMB observations constrain the dark energy density at
decoupling to be less then 30 per cent of the critical one
\cite{BEAN2}. Such a non negligible contribution can be realized in
a large class of models and therefore cannot be {\em a priori}
excluded. Consequently it is of crucial importance to find an
appropriate parametrization for the dark energy equation of state
that allows us to take into account the full impact dark energy
has on different types of cosmological observations. In this
paper we attempt to address this very issue by providing a formula
to describe $w_Q$, that can accommodate most of the proposed dark
energy models in terms of physically motivated parameters. We will
then be in a position to benefit from the fact that the
cosmological constraints on these parameters will allow us to
infer general information about the properties of dark energy.\\

\section{Dark energy equation of state}

The dynamics of the quintessence field $Q$ for a given potential $V$
is described by the system of equations:
\begin{equation}
\ddot{Q}+3H\dot{Q}+\frac{dV}{dQ}=0,
\label{klein}
\end{equation}
and
\begin{equation}
H^2=\frac{8\pi G}{3}\left[\rho_{m}+\rho_{r}+\frac{\dot{Q}^2}{2}+V(Q)\right],
\label{friedmann}
\end{equation}
where $\rho_{m}$ and $\rho_{r}$ are the matter and radiation
energy densities respectively and the dot is the time derivative.
The specific evolution of $w_Q(a)$, where
$a$ is the scale factor, depends on the shape of the potential,
however there are some common features in its behaviour that
can be described in a model independent manner and which allow us
to introduce some physical parameters. As a first
approach, we notice that a large number of quintessence models are
characterized by the existence of the so called 'tracker regime'.
It consists of a period during which the scalar field, while it is
approaching a late time attractor, evolves with an almost constant
equation of state whose value can track that of the background
component. The necessary conditions for the existence of tracker
solutions arising from scalar field potentials
has been studied in \cite{CALD,STEIN}. In this paper, we consider
a broad class of tracking potentials.
 These include models for which $w_Q(a)$ evolves
smoothly, as with the inverse power law \cite{ZLATEV}, $V(Q) \sim
1/Q^{\alpha}$ (INV) and the supergravity inspired potential
\cite{BRAX}, $V(Q) \sim 1/Q^{\alpha} e^{Q^2/2}$ (SUGRA).
Late time rapidly varying equation of states arise in potentials with two
exponential functions \cite{BARRE}, $V \sim e^{- \alpha
Q}+e^{\beta Q}$ (2EXP), in the so called `Albrecht \& Skordis'
model \cite{ALB} (AS) and in the model proposed by Copeland et al.
\cite{COPE} (CNR). To show this in more detail, in fig.1 we plot
the equation of state obtained by solving numerically
Eq.~(\ref{klein}) and Eq.~(\ref{friedmann}) for each of these
potentials.

\begin{figure}[ht]
\includegraphics[height=8cm,width=8cm]{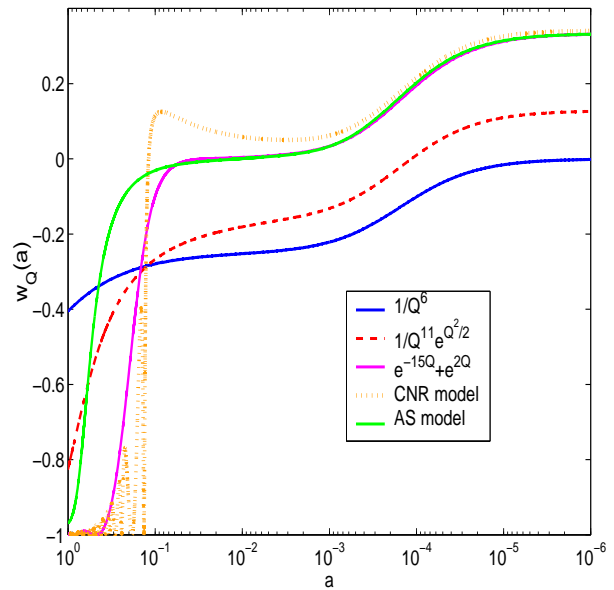}
\caption{\label{fig:eqs}Evolution of $w_{Q}$ against the scale factor
for an inverse power law model (solid blue line),
SUGRA model (\cite{BRAX}) (dash red line),
two exponential potential model (\cite{BARRE}) (solid magenta line),
AS model (\cite{ALB}) (solid green line) and
CNR model (\cite{COPE}) (dot orange line).}
\end{figure}
\begin{figure}[ht]
\includegraphics[height=8cm,width=8cm]{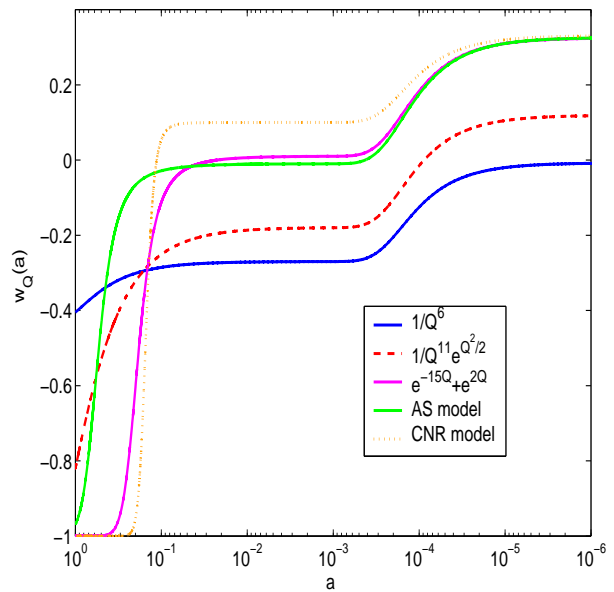}
\caption{\label{fig:best} Plot of $w_Q^{p}(a)$ best fit
for different potentials}
\end{figure}

There are some generic features that appear to be present, and
which we can make use of in our attempts to parameterize $w_Q$.
For a large range of initial conditions of the quintessence field,
the tracking phase starts before matter-radiation equality. In
such a scenario $w_Q(a)$ has three distinct phases, separated by
two `phase transitions'. Deep into both the radiation and matter
dominated eras the equation of state, $w_Q(a)$, takes on the
values $w_Q^r$ and $w_Q^m$ respectively, values that are related
to the equation of state of the background component $w_B$ through
\cite{CALD,STEIN}:

\begin{equation}
\ w_Q \approx \frac{w_B-2(\Gamma-1)}{2\Gamma-1},
\label{track}
\end{equation}
where $\Gamma=V''V/(V')^2$ and $V' \equiv dV/dQ$ etc.
For the case of an exponential potential, $\Gamma=1$,
with $w_Q=w_B$, but
in general $w_Q \neq w_B$. Therefore if we do not specify
the quintessence potential the values of
$w_Q^r$ and $w_Q^m$ should be considered as free parameters.

The two transition phases can each be described by two parameters;
the value of the scale factor $a_c^{r,m}$ when the equation of state $w_Q$
begins to change and the width $\Delta_{r,m}$ of the transition.
Since $\Gamma$ is constant or slowly varying during the tracker regime,
the transition from $w_Q^r$ to $w_Q^m$ is always
smooth and is similar for all the models (see fig.~1).
To be more precise, we have found
that $a_c^r\sim 10^{-5}$ and $\Delta_r\sim 10^{-4}$ during this transition,
the former number expected from the time of matter-radiation
equality and the latter from the transition period from radiation to
matter domination.
However, when considering the transition in $w_Q$ from $w_Q^m$ to
the present day value $w_Q^0$, we see from fig.~1 that this can
be slow ($0 < a_c^{m}/ \Delta_m < 1$)
or rapid ($a_c^{m}/ \Delta_m > 1$) according to the slope of
the quintessence potential. For instance in models with a steep
slope followed by
a flat region or by a minimum, as in the case of the two exponentials,
the AS potential
or the CNR model, the scalar field evolves towards a solution that
approaches the late time attractor, finally deviating from the tracking
regime with the parameter $\Gamma$ rapidly varying.
In contrast the inverse power law potential always has a
slower transition since $\Gamma$ is constant for all times.
Given these general features we conclude that the behavior of $w_Q(a)$ can be
explained in terms of functions, $w_Q^p(a)$, involving the following parameters:
$w_Q^0$, $w_Q^m$, $w_Q^r$, $a_c^{m}$ and $\Delta_m$.
The authors of \cite{BASS} have recently used an
expansion in terms of a Fermi-Dirac function in order to constrain a class of
dark energy models with rapid late time
transitions. In what follows we find that a generalisation of this involving
a linear combination of such functions allows for a wider range of models to be
investigated. To be more precise, we propose the following formula for $w_Q^p(a)$:

\begin{equation}
w_Q^{p}(a)=F_1f_r(a)+F_2f_m(a)+F_3,
\label{exp}
\end{equation}
with
\begin{equation}
f_{r,m}(a)=\frac{1}{1+e^{-\frac{a-a_c^{r,m}}{\Delta_{r,m}}}}.
\end{equation}

The coefficients $F_1$, $F_2$ and $F_3$ are determined by
demanding that $w_Q^{p}(a)$ takes on the respective values
$w_Q^r$, $w_Q^m$, $w_Q^0$ during radiation ($a_r$) and matter
($a_m$) domination as well as today ($a_0$). Solving the algebraic
equations that follow we have:
\begin{widetext}
\begin{equation}
F_1=\frac{\left( w_Q^m-w_Q^r \right) \left( f_m(a_0)-f_m(a_r)
\right)-
 \left( w_Q^0-w_Q^r \right) \left( f_m(a_m)-f_m(a_r) \right)}
 {\left( f_r(a_m)-f_r(a_r) \right) \left( f_m(a_0)-f_m(a_r) \right)-
\left( f_r(a_0)-f_r(a_r) \right) \left( f_m(a_m)-f_m(a_r)
\right)},
\end{equation}

\begin{equation}
F_2=\frac{ w_Q^0-w_Q^r }{ f_m(a_0)-f_m(a_r)}-F_1
\frac{f_r(a_0)-f_r(a_r)}{f_m(a_0)-f_m(a_r)},
\end{equation}

\begin{equation}
F_3=w_Q^r-F_1f_r(a_r)-F_2f_m(a_r),
\end{equation}
\end{widetext}
where $a_0=1$, and the value of and $a_r$ and $a_m$ can be
arbitrarily chosen in the radiation and matter era because of the
almost constant nature of $w_Q$ during those eras. For example in
our simulations we assumed $a_r=10^{-5}$ and $a_m= 10^{-3}$. In
table 1 we present the best fit parameters obtained minimizing a
chi-square for the different models of fig.1, and in fig.2 we plot
the associated functions $w_Q^{p}(a)$. It is encouraging to see
how accurately the Fermi-Dirac functions mimic the exact time
behavior of $w_Q(a)$ for the majority of the potentials. In fig.3
we plot the absolute value of the difference $\Delta w(a)$ between
$w_Q(a)$ and $w_Q^{p}(a)$. The discrepancy is less then $1\%$ for
redshifts $z<10$ where the energy density of the dark energy can
produce observable effects in these class of models and it remains
below $9\%$ between decoupling and matter-radiation equality. Only
the CNR case is not accurately described by $w_Q^{p}(a)$ due to
the high frequency oscillations of the scalar field which occur at
low redshift as it fluctuates around the minimum of its potential.
In fact these oscillations are not detectable, rather it is the
time-average of $w_Q(a)$ which is seen in the cosmological
observables, and can be described by the corresponding
$w_Q^{p}(a)$. There are a number of impressive features that can
be associated with the use of $w_Q^p(a)$ in Eq.~(\ref{exp}).
For instance it can reproduce not only the behavior of models
characterized by the tracker regime but also more general ones. As
an example of this in fig.4 we plot $w_Q^{p}(a)$ corresponding to
three cases: a K-essence model \cite{MUKA} (blue solid line); a
rapid late time transition \cite{PARKER} (red dash-dot line) and
finally one with an equation of state $w_Q^0 < -1$ (green dash
line). The the observational constraints on $w_Q^0$, $w_Q^m$,
$w_Q^r$, $a_c^{m}$ and $\Delta_m$ lead to constraints on a large
number of dark energy models, but at the same time it provides us
with model independent information on the evolution of the dark
energy. It could be argued that the five dimensional parameter
space we have introduced is too large to be reliably constrained.
Fortunately this can be further reduced without losing any of the
essential details arising out of tracker solutions in these
Quintessence models. In fact nucleosynthesis places tight
constraints on the allowed energy density of any dark energy
component, generally forcing them to be negligible in the
radiation era \cite{MEL,YA}. The real impact of dark energy occurs
after matter-radiation equality, so we can set $w_Q^r=w_Q^m$.
Consequently we end up with four parameters: $w_Q^0$, $w_Q^m$,
$a_c^{m}$ and $\Delta_m$. Although they increase the already large
parameter space of cosmology, they are necessary if we are to
answer fundamental questions about the nature of the dark energy.
The parameters make sense, if $w_Q(a)$ evolves in time, we need to
know when it changed ($a_c^{m}$), how rapidly ($\Delta_m$) and
what its value was when it changed ($w_Q^m$). Neglecting the
effects during the radiation dominated era it proves useful to
provide a shorter version of Eq.~(\ref{exp}), in fact since we can
neglect the transition from radiation to matter dominated eras,
then the linear combination Eq.~(\ref{exp}) can be rewritten as
\footnote[1]{We thank Eric Linder for pointing this out to us.}:

\begin{equation}
w_Q^{p}(a)=w_Q^0+(w_Q^m-w_Q^0)\times\frac{1+e^\frac{a_c^m}{\Delta_m}}
{1+e^{-\frac{a-a_c^m}{\Delta_m}}}
\times\frac{1-e^{-\frac{a-1}{\Delta_m}}}{1-e^\frac{1}{\Delta_m}}.
\label{lowz}
\end{equation}

As we can see in fig.5, the relative difference between the exact
solution $w_Q(a)$ of the Klein-Gordon equation and
Eq.~(\ref{lowz}) is smaller than $5\%$ for redshifts $z<1000$,
therefore it provides a very good approximation for the evolution
of the quintessence equation of state. Both Eq.~(\ref{exp})
and Eq.~(\ref{lowz}) are very useful in that they allow us to take
into account the clustering properties of dark energy (see for
instance \cite{DAVE}) and to combine low redshift measurements
with large scale structure and CMB data. We would like to point
out that one of the key results in this paper is the fact that our
expansion for the equation of state allows for a broader range of
redshift experiments than has previously been proposed.

\begin{table}
\caption{\label{tab:table1} Best fit values of the parameters of the
expansion ~(\ref{exp}).}
\begin{ruledtabular}
\begin{tabular}{cccccccc}
 &$w_Q^0$ &$w_Q^m$ &$w_Q^r$& $a_c^m$ &$\Delta_m$\\
\hline
INV& -0.40 & -0.27 & -0.02 & 0.18 & 0.5 \\
SUGRA& -0.82 & -0.18 & 0.10 & 0.1 & 0.7 \\
2EXP& -1 & 0.01 & 0.31 & 0.19 & 0.043 \\
AS& -0.96 & -0.01 & 0.31 & 0.53 & 0.13 \\
CNR& -1.0 & 0.1 & 0.32 & 0.15 & 0.016 \\
\end{tabular}
\end{ruledtabular}
\end{table}

\begin{figure}[ht]
\includegraphics[height=8cm,width=8cm]{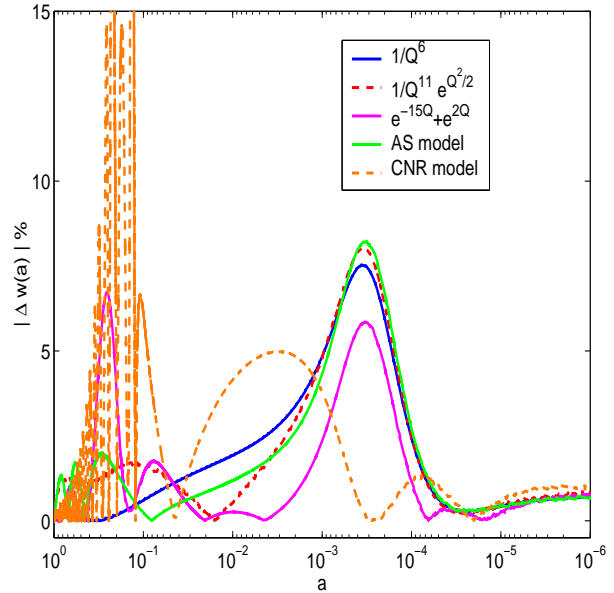}
\caption{\label{fig:werr} Absolute value of the difference
 between $w_Q(a)$ and $w_Q^{p}(a)$ for the models of fig.1.}
\end{figure}

\begin{figure}[ht]
\includegraphics[height=8cm,width=8cm]{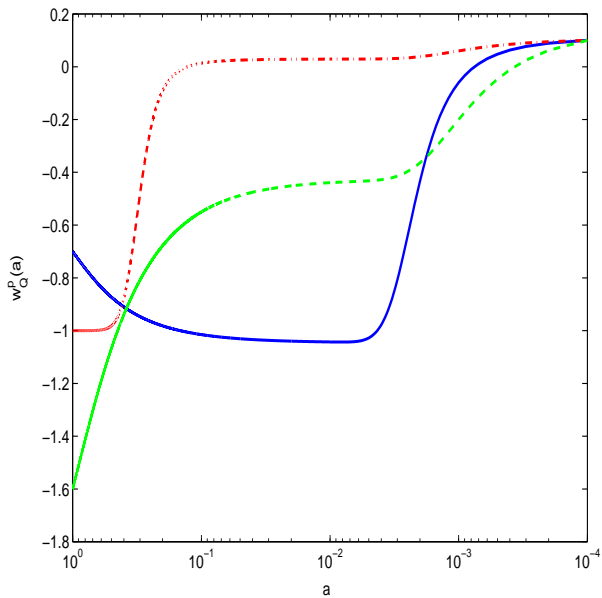}
\caption{\label{fig:altern} Time evolution of $w_Q^p(a)$ as in the case
of K-essence (blue solid line), late time transition (red dash-dot line)
end with $w_Q^o<-1$ (green dash line).}
\end{figure}

\begin{figure}[ht]
\includegraphics[height=8cm,width=8cm]{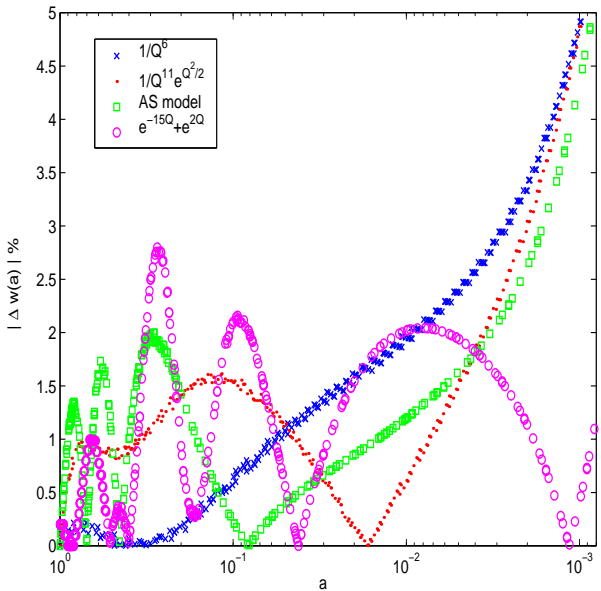}
\caption{\label{fig:errd} Absolute value of the difference
 between $w_Q(a)$ and the low redshift formula Eq.~(\ref{lowz})
 for the models of fig.1.}
\end{figure}

\section{Conclusions}

The evidence for a present day accelerating universe appears to be
mounting, and accompanied with it is the need to understand the
nature of the dark energy that many believe to be responsible for
this phenomenon. Of the two possibilities proposed to date, a
cosmological constant or a Quintessence scalar field, the latter
suffers in that there are a plethora of models that have been
proposed, all of which satisfy the late time features, that of an
accelerating universe. Yet there is no definitive particle physics
inspired model for the dark energy therefore an other route should be
explored, trying to determine the equation
of state that the dark energy satisfies in a model independent
manner. If this were to work it would allow us to discuss the
impact dark energy has had on cosmology without the need to refer
to particular dark energy scenario. In this paper, we have
begun addressing this approach.
We have introduced a parameterization of the dark energy equation
of state $w_Q^p(a)$ which involves five parameters and shown how
well they reproduce a wide range of dark energy models.
By estimating their value
from cosmological data we can constrain the dark energy in a
model independent way.
This could be important in future years
when high precision CMB and large scale structure observations
begin to probe medium to large redshifts, regions where differences
in the features of the Quintessence models begin to emerge.
Although the low redshift data indicate that it is currently
impossible to discriminate a cosmological constant from a
Quintessence model, moving to higher redshifts may result in
evidence for a time varying equation of state in which case,
we need to be in a position to determine such
an equation of state in a model independent manner. We believe
this paper helps in this goal.

\acknowledgments
We would like to thank Carlo Ungarelli, Nelson J. Nunes, Michael Malquarti
and Sam Leach for useful discussions.
In particular we are grateful to Eric Linder for
the useful suggestions he made. PSC is supported
by a University of Sussex bursary.

\end{document}